\input{aipcheck}

\documentclass[
    ,final            
  ]
  {aipproc}
\layoutstyle{8x11single}

\usepackage{bm}%
\usepackage{amssymb}%
\newcommand{\foot}[2]{\mbox{\tiny $\stackrel{#1}{#2}$}}
\newcommand{\el}{\ \\\nonumber}

\newcommand{\sh}[1]{#1\hskip-6pt \diagup}

\newcommand{\ket}[1]{\left|#1\right>}

\newcommand{\bracket}[3]{\left<#1\left|#2\right|#3\right>}

\newcommand{\mb}[1]{\bm{#1}}

\newcommand{\dpt}{(2\pi)^3}

\newcommand{\gfiv}{\gamma^5}

\newcommand{\eg}{{\it e.~g. }}
 

\usepackage{epstopdf}

\begin{document}

\title{Electroweak Single Pion Production and Form Factors of the $\Delta(1232)$ Resonance}

\classification{13.15.+g, 13.60.Le}
\keywords      {Pion, electron, neutrino, deuteron}

\author{Jakub \.{Z}muda, Krzysztof M. Graczyk}{
  address={Institute for Theoretical Physics, University of Wroc\l aw, pl. M. Borna 9,
50-204, Wroc\l aw, Poland}
}

\begin{abstract}
We extend and review our analysis of the nucleon $\to \Delta(1232)$ transition electroweak form factors from Ref. \cite{Graczyk:2014dpa}.
New fit of the $\Delta(1232)$ vector form factors to electron-proton scattering $F_2$ structure function is introduced as well, leading to results different from the popular parametrization of Ref. \cite{Lalakulich:2006sw}. A clear model dependence of the extracted parameters emerges. Fit to neutrino scattering data is performed in all available isospin channels. The resulting axial mass is $M_{A\Delta}=0.85\foot{+0.09}{-0.08}\;$(GeV) and $C_5^A(0) = 1.10\foot{+0.15}{-0.14}$. The latter value is in accordance with Goldberger-Treiman relation as long as the deuteron effects are included.
\end{abstract}

\maketitle

\section{Introduction}
\label{sec:spp}

The problem of single pion neutrinoproduction (SPP) has been studied for many decades.
Its importance has become clear with the development of neutrino accelerator experiments,
such as MINOS \cite{Evans:2013pka}, T2K \cite{Abe:2011ks}, NOvA \cite{Ayres:2004js}, 
MiniBooNE \cite{AguilarArevalo:2007it}, and LBNE \cite{Adams:2013qkq}.
In the few-GeV energy region chartacteristic for the above mentioned experiments
this interaction channel contributes a large fraction of the total cross section.
One estimates, that for an isoscalar target and neutrino energy of
around 1 GeV SPP accounts for about 1/3 of the interactions. 

These SPP events give rise to the background in measurements of quasi-elastic neutrino scattering off nuclear targets if
subsequent pion absorption occurs. For experiments aiming at electron neutrino appearance measurement neutral current  $\pi^0$ production process
adds to the background in water Cherenkov detectors. Correct understanding and modeling of the cross-sections for the SPP is crucial
for precise extraction of neutrino oscillation parameters in long baseline experiments.

Theoretical modelling of the SPP processes on nuclear targets is biased by systematic errors coming from nuclear model
uncertainties. They are driven by the strong nature of hadron interactions inside the nucleus, which do not allow
for a feasable, exact solution of the problem.
Experimental measurements also suffer from these efects. An apparent tension between the
MiniBooNE and very recent MINER$\nu$A SPP data on (mostly) carbon target (Ref. \cite{Eberly:2014mra,nasza2014}) is one of the key examples.
For the purpose of analysis of Nucleon (N) to $\Delta$(1232) resonance transition vertex one desires measurements of the neutrino-production on free or almost free targets.
Such data exist only for $\sim$30 years old Argonne National Laboratory (ANL) \cite{Barish:1978pj,Radecky:1981fn} and Brookhaven National Laboratory (BNL) 
\cite{Kitagaki:1986ct,Kitagaki:1990vs} bubble chamber experiments, where deuteron and hydrogen targets were utilized.
In this case one may hope to reduce the many-body bias in a reasonable manner with a simple theoretical ansatz \cite{AlvarezRuso:1998hi}.

One can not understand the neutrino SPP data without introducing an appropriate nonresonant background,
see Ref. \cite{Fogli:1979cz}. More recent studies of weak SPP fit the $N\to\Delta$ transition axial form factors utilizing
only the neutrino-proton channel $\nu_\mu + p\to  \mu^-+\pi^++p$ \cite{Sato:2003rq,Hernandez:2007qq,Barbero:2008zza,Hernandez:2010bx,Lalakulich:2010ss,Serot:2012rd}. Simple total
cross section ratio analysis shows, that the background contribution is much larger in neutrino-neutron channels. The neutrino-proton SPP 
channel can be described well within a model that contains the $\Delta(1232)$ resonance
contribution only, see \eg Ref. \cite{Graczyk:2009qm}. Thus a quantitative, statistical, validation of any pion neutrinoproduction model
should be done using all available isospin channels, where one gets more information about the background contribution.
In Ref. \cite{Graczyk:2009qm} a consistent fit for both ANL and BNL
data sets with deuteron effects included yielded $C_ 5^A(0)=1.19\pm 0.08$ and $M_A=0.94 \pm 0.03$ GeV. The attempt to extract the leading $C_5^A(Q^2)$
$N\to\Delta$ form factor parameters in a model containing both nonresonant background and deuteron effects has been done in Ref. \cite{Hernandez:2010bx}.
The results gave the value of $C_5^A(0)=1.00\pm 0.11$ far from the Goldberger-Treiman relation estimate of $C_5^A(0)\approx 1.15$ \cite{Goldberger:1958tr,BarquillaCano:2007yk}.
Lack of the deuteron effects pushes the fit even further away from theoretical prediction, giving $C_5^A(0)=0.867\pm 0.075$ in Ref. \cite{Hernandez:2007qq}.
From the above mentioned models only those in Refs. \cite{Sato:2003rq,Barbero:2008zza} have been directly 
tested against the electroproduction data. Authors of \cite{Hernandez:2007qq,Hernandez:2010bx,Lalakulich:2010ss} use vector form factor parametrization
from Ref. \cite{Lalakulich:2006sw}, based on the MAID analysis \cite{Drechsel:2007if}. The $\Delta$ form factors have been fitted directly
to the $\Delta$ helicity amplitudes from MAID. Thus the approach from Ref. \cite{Lalakulich:2006sw} relied only on the $\Delta$ resonance excitation without any nonresonant background. The problem is that the resonance helicity amplitudes extraction procedure
is model-dependent. One has to make some assumptions how to separate $\Delta$ and background contributions from the data
and $\Delta$ -- background interference effects are strong. We show later that if one uses form-factors obtained within one description
of pion electroproduction in a model with different physical components the resulting cross sections may become imprecise.

The above mentioned caveats of previous analyses have motivated us to propose an improved approach. We adapt and develop the statistical framework of
Ref. \cite{Graczyk:2009qm} in order to
fit both vector and axial form factors of the $\Delta(1232)$ resonance. We use inclusive electron-proton scattering data for the electromagnetic interaction in the $\Delta(1232)$ region, adding a new fit of the form factors from Ref. \cite{Lalakulich:2006sw}. The deuteron bubble chamber data of ANL and BNL experiments are used for the weak interaction. For this analysis we are the first authors to incorporate the neutron channels.
In this manner we include the data sets, that are very sensitive to the nonresonant background.

\section{Formalism}
\label{sec:genform}

In the neutrino single pion production off free nucleon targets one distinguishes three isospin channels:
\begin{eqnarray}
\label{channel_1}
\nu_\mu(l) + p(p) & \to & \mu^-(l') + \pi^+(k) + p(p')
\\
\label{channel_2}
\nu_\mu(l) + n(p) & \to & \mu^-(l') + \pi^0(k) + p(p')
\\
\label{channel_3}
\nu_\mu(l) + n(p) & \to & \mu^-(l') + \pi^+(k) + n(p')
\end{eqnarray}
with $l$, $l'$, $p$, $p'$ and $k$ being the neutrino, muon, initial nucleon, final nucleon and pion four momenta respectively. 
The definition of four momentum transfer is following:
\begin{eqnarray}
\label{eq:4momtransf}
q=l-l'=p'+k-p,\quad Q^2=-q^2,\quad q^\mu = (q^0,\mathbf{q})
\end{eqnarray}
and the square of hadronic invariant mass is:
\begin{eqnarray}
\label{eq:W}
W^2=(p+q)^2=(p'+k)^2.
\end{eqnarray}
Metric convention $g^{\mu\nu}=\mathrm{diag(+,-,-,-)}$ is used throughout this paper.

For the pion electroproduction we are interested in proton target reactions;
\begin{eqnarray}
\label{channel_1e}
e^-(l) + p(p) & \to & e^-(l') + \pi^+(k) + n(p')
\\
\label{channel_2e}
e^-(l) + p(p) & \to & e^-(l') + \pi^0(k) + p(p').
\end{eqnarray}
The proton interaction channel (\ref{channel_1}) is dominated by the intermediate $\Delta^{++}$ resonance
excitation, which makes it very sensitive to the properties of this resonance. Neutron channels (Eqs. (\ref{channel_2}) and (\ref{channel_3}))
contain a large contribution of nonresonant pion production, thus they present more challenges for theorists. They are crucial
to verification of any consistent SPP model.

\subsection{$N\to\Delta(1232)$ transition}
\label{sec:delta}

We treat the $\Delta(1232)$ resonance excitation within the isobar framework.
The most general form of positive parity spin-$\frac{3}{2}$ particles electroweak excitation vertex can be expressed as:
\begin{equation}
\nonumber
\label{eq:delver}
\Gamma^{\alpha\mu}(p,q)
=
\left[V^{\alpha\mu}_{3/2}+ A^{\alpha\mu}_{3/2}\right]\gfiv
\end{equation}
where
 \begin{eqnarray}
 V^{\alpha\mu}_{3/2} &=& 
  \frac{C^V_3(Q^2)}{M}(g^{\alpha\mu}\sh{q} \hspace{-1pt} - \hspace{-1pt}
q^\alpha \gamma^\mu ) \hspace{-1pt}  + \hspace{-1pt}  \frac{C^V_4(Q^2)}{M^2}(g^{\alpha\mu}q  \hspace{-1pt}\cdot \hspace{-1pt} (p\hspace{-1pt}  + \hspace{-1pt} q) \hspace{-1pt} -  \hspace{-1pt}q^\alpha (p\hspace{-1pt}  +  \hspace{-1pt} q)^\mu)\hspace{-1pt}  + \hspace{-1pt}
\frac{C^V_5(Q^2)}{M^2}(g^{\alpha\mu}q \hspace{-1pt} \cdot  \hspace{-1pt} p \hspace{-1pt} -  \hspace{-1pt} q^\alpha p^\mu ) \hspace{-1pt} +  \hspace{-1pt}g^{\alpha\mu} C_6^V(Q^2)  \nonumber \\ 
\\
A^{\alpha\mu}_{3/2} 
& = & 
\left[ \frac{C^A_3(Q^2)}{M}(g^{\alpha\mu}\sh{q}  \hspace{-1pt} -  \hspace{-1pt} q^\alpha\gamma^\mu )  \hspace{-1pt}+ \hspace{-1pt} \frac{C^A_4(Q^2)}{M^2}(g^{\alpha\mu}q  
\hspace{-1pt} \cdot\hspace{-1pt}  (p  +  q)\hspace{-1pt}  -\hspace{-1pt}
q^\alpha (p \hspace{-1pt} +\hspace{-1pt}  q)^\mu) \hspace{-1pt}+\hspace{-1pt}  C_5^A(Q^2) g^{\alpha\mu} \hspace{-1pt} +\hspace{-1pt}  \frac{C^A_6(Q^2)}{M^2}q^\alpha q^\mu\right] \gfiv.
\end{eqnarray}

A relevant information about the inner structure of the $\Delta(1232)$ resonance is contained in a set of vector and axial form factors
$C_j^{V,A}$. In this paper they are assumed to be functions of $Q^2$ only (with the exception of $C_4^V$ which depends also on $W$).

\subsubsection{Vector contribution}

The conserved vector current (CVC) hypothesis gives the relation between weak and electromagnetic vector form factors.
In the case of $\Delta$(1232) resonance and hereby used convention both sets are exactly the same.
The size and excellent accuracy of the electromagnetic data set allows for an introduction of multiple fit parameters.

We explore two models of vector form factors. The first parametrization, refered to as ``Model I'', has the same functional form,
as in Ref. \cite{Lalakulich:2006sw}:
\begin{eqnarray}\label{eq:modela}
C_3^V(Q^2)&=&\frac{C_3^V(0)}{\displaystyle \left(1+ D \cdot \frac{Q^2}{M_v^2}\right)^2}\frac{1}{\displaystyle 1+ A\, \frac{Q^2}{M_v^2}}\\
C_4^V(Q^2)&=&\frac{C_4^V(0)}{\displaystyle \left(1+ D \cdot \,\frac{Q^2}{M_v^2}\right)^2}\frac{1}{\displaystyle 1+A \, \frac{Q^2}{M_v^2}}\\\label{eq:enda}
C_5^V(Q^2)&=&\frac{C_5^V(0)}{\displaystyle \left(1+ D \cdot \frac{Q^2}{M_v^2}\right)^2}\frac{1}{\displaystyle 1+ B\, \frac{Q^2}{M_v^2}}.
\end{eqnarray}
In the above equations $M_V=0.84\;$GeV is the standard vector mass. Everything else is treated as a fit parameter.

We also propose our own model of electromagnetic form factors.
We assume that the $N\to \Delta$ transition form factors have the same large $Q^2$  behaviour
as the electromagnetic elastic nucleon form factors. There exist theoretical arguments 
 \cite{Brodsky:1974vy} suggesting that at $Q^2 \to \infty $ the nucleon form factors fall down as $1/Q^4$. Following these assumptions we adopt
appropriate Pad\'{e} type parametrization used previously to parametrize the electromagnetic form factors of the nucleon \cite{Kelly:2004hm}.
In this manner we allow for a  deviation from the $SU(6)$-symmetry quark model relations $C_4^V(Q^2)=-({M}/{W})C_3^V(Q^2)$ 
and $C_5^V=0$ between the form factors \cite{Liu:1995bu}. Finally, we assume  the  dipole representation of $C_5^V(Q^2)$ to reduce the number of parameters.
Altogether, our parametrization has the following form:
\begin{eqnarray}
\label{eq:Kellyc5}
C_3^V(Q^2)&=&\frac{C_3^V(0)}{1+AQ^2+BQ^4+CQ^6}\cdot(1+K_1Q^2)\\
C_4^V(Q^2)&=&-\frac{M_p}{W}C_3^V(Q^2)\cdot\frac{1+K_2Q^2}{1+K_1Q^2}\\
\label{eq:c5K} 
C_5^V(Q^2)
&=&
\frac{C_5^V(0)}{\displaystyle \left(1+D\, \frac{Q^2}{M_V^2}\right)^2}.
\end{eqnarray}
This parametrization reproduces quark model relation between $C_3^V$ and $C_4^V$ at $Q^2=0$. It is also allows for nonzero value of $S_{1/2}$ helicity
amplitude. We call it ``Model II''.

\subsubsection{Axial contribution}

Here the leading contribution comes from $C_5^A(Q^2)$ which is an analogue of the isovector nucleon axial form factor.
Partially conserved axial current (PCAC) hypothesis relates the value of $C_5^A(0)$
with the strong  coupling constant $f^\ast$ through off-diagonal Goldberger-Treiman relation \cite{Goldberger:1958tr,BarquillaCano:2007yk}:
\begin{eqnarray}
\label{eq:GolTreq}
C_5^A(0)=\frac{f^\ast}{\sqrt{2}}\approx 1.15,
\end{eqnarray}
but we will treat $C_5^A(0)$ as a free parameter.
We assume, that $C_5^A$ has a dipole $Q^2$ dependence:
\begin{eqnarray}
\label{eq:dipolec5a}
C_5^A(Q^2)=\frac{C^A_5(0)}{\displaystyle \left(1+\frac{Q^2}{M_{ A\Delta}^2}\right)^2}
\end{eqnarray}
The axial mass parameter $M_{A\Delta}$ can be related to ``resonance axial charge radius''.
It is also subject to fit, but we expect it to be of the order of $1\, \mathrm{ GeV}$. 

The $C_6^A$ form factor is an analogue of the nucleon induced pseudoscalar form factor. One can use PCAC to relate it to $C_5^A$:
\begin{eqnarray}
\label{eq:c6a}
C_6^A(Q^2)=\frac{M^2}{\displaystyle m_\pi^2+Q^2}\, C_5^A(Q^2),
\end{eqnarray}
where $m_\pi$ is average pion mass.
The $C_3^A(Q^2)$ is the axial counterpart of the very small electric quadrupole (E2) transition form factor. Unfortunately,
bubble chamber data set is too inacurate to precisely measure its effect. Due to expected similarities between $\Delta$ and nucleon
properties we set $C_3^A=0$. For the $C_4^A$ we use the Adler model relation \cite{Adler}:
\begin{eqnarray}
\label{eq:c4a}
C_4^A(Q^2)=-C_5^A(Q^2)/4.
\end{eqnarray}
In this way the axial contribution is fully determined by $C_5^A(Q^2)$.
Altogether there are two free parameters: $C_5^A(0)$ and $M_{A\Delta}$.
If there were enough experimental data one could drop the Adler relation and treat $C_4^A(Q^2)$ as an independent form factor or even
determine, whether $C_3^A(Q^2)$ has nonzero value. 
However, the  ANL and BNL experimental data do not have sufficient statistics even to obtain separate fits of $C_5^A$ and $C_4^A$ \cite{Graczyk:2009zh}, 
see also the discussion in Ref. \cite{Hernandez:2010bx}.  

\subsection{Cross section}
\begin{figure}[!htb]
\centering\includegraphics[width=0.8\columnwidth]{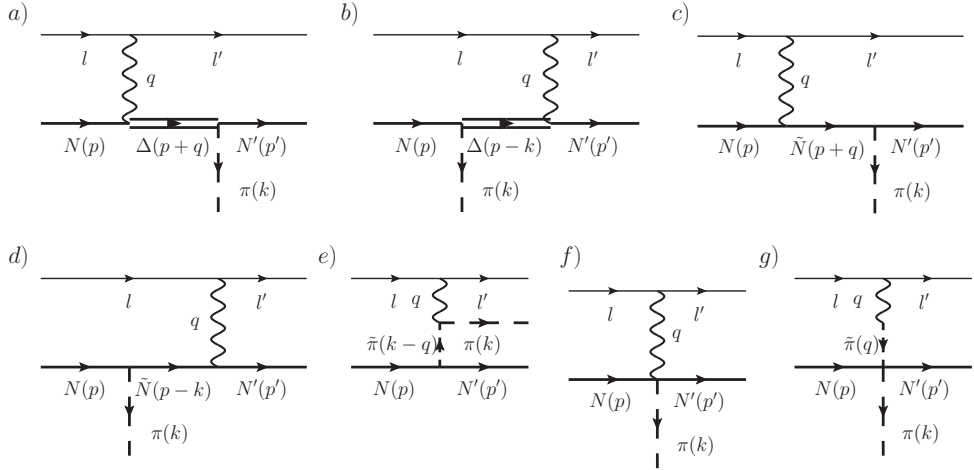}
\caption{(Color online) Basic pion production diagrams from \cite{Hernandez:2007qq}:
a) Delta pole ($\Delta$P), b) crossed  Delta pole (C$\Delta$P), c) nucleon pole (NP), d) crossed nucleon pole (CNP),
e) pion-in-flight (PIF), f) contact term (CT) and g) pion pole (PP).}\label{fig:mec}
\end{figure}

We express the inclusive double differential SPP cross section for neutrino scattering off nucleons at rest as:
\begin{eqnarray}\nonumber\label{eq:3diffnucleon}
\frac{d^2\sigma}{dQ^2dW}&=&\frac{1}{32\pi E^2} G^2_F\cos^2\theta_C \frac{W}{M^2} \int \frac{d^3k}{\dpt 2E_\pi (k) E(p')}L_{\mu\nu}A^{\mu\nu}\delta ( E(p')+ E_\pi (k) - M- q^0)\el
L^{\mu\nu}&=&l^\mu l'^{\nu}+l^\nu l'^{\mu}-g^{\mu\nu}l\cdot l'+i\epsilon^{\mu\nu\alpha\beta}l'_{\alpha}l_{\beta}\\
A^{\mu\nu}&=&\sum_{spins}\bracket{\pi N'}{j^\mu_{cc}(0)}{N}\bracket{\pi N'}{j^\nu_{cc}(0)}{N}^\ast.
\end{eqnarray}
where $E$ is the incident neutrino energy, $M$ is the averaged nucleon mass, $E_\pi(k)$ and $E(p')$ are the final state pion and nucleon energies, 
$G_F=1.1664\cdot 10^{-11}\;$MeV$\mathrm{^{-2}}$ is the Fermi constant, $L^{\mu\nu}$ - the leptonic and $A^{\mu\nu}$ - the hadronic tensors. 
The Cabibbo angle, $\cos(\theta_C)=0.974$, was factored out of the weak charged current definition.

The information about dynamics of SPP is contained in transition matrix elements, $\bracket{\pi N'}{j^\mu_{cc}(0)}{N}$, 
between an initial nucleon state $\ket{N}$ and a final nucleon-pion state $\ket{\pi N'}$.

In the model of this paper the dynamics of SPP process is defined by a set of Feynman diagrams (Fig. \ref{fig:mec})
with vertices determined by the effective chiral field theory. They are discussed in Ref. \cite{Hernandez:2007qq}, 
where one can find exact expressions for $j^\mu$.
The same set of diagrams describes also pion electroproduction, with the exception of the pion pole diagram,
which is purely axial. We call this approach "HNV model" after the names of the authors of Ref. \cite{Hernandez:2007qq}.

\subsubsection{Deuteron effects}

In this paper we consider a deuteron model based on phenomenological nucleon momentum distribution, $f(p)$, taken from the Paris potential \cite{Lacombe:1981eg}.
It is assumed that the spectator nucleon does not participate in the interaction and that there are no final state interactions (FSI). This assumption
is based on the results of Ref. \cite{Shen:2012xz} where for the quasielastic scattering case
FSI were shown to be negiligible as long as the neutrino energy is larger than $500$~MeV. The very recent study (newer, than the hereby analysis)
of Ref. \cite{Wu:2014rga} proved the FSI to be important in the $n\pi^+$ channel, where they lead to
a substantial reduction of the cross section for forward-going pions. Further studies regarding the impact of FSI effects are needed,
also in the ANL and BNL experimental data analysis where the event selection is based on spectator approach as well.
We introduce also the effective binding energy:
\begin{equation}
B(p) =2 E(p) - M_D,
\end{equation}
where $M_D$ is deuteron mass. The expression for the cross section becomes:
\begin{eqnarray}
\label{eq:dq2deut}
\frac{d\sigma}{dQ^2 dW}= \int d^3p
\frac{f(p)}{v_{rel.}} \frac{G_F^2\cos^2(\Theta_C)|\mb{l'}|}{16\pi E_\nu E(p)|\mathcal{J}|}\int\frac{d^3k}
{\dpt 2E_\pi(k)}\int\frac{d^3p'}{\dpt 2E(p')} L_{\mu\nu}A^{\mu\nu}(p,\tilde{q},k)\delta^4(p+\tilde{q}-k-p').
\end{eqnarray}
with $\tilde q^\mu = (q^0-B(p), \vec q)$ and $v_{rel.}=\sqrt{(l\cdot p)^2}/EE(p)$.
We also define a Jacobian $\mathcal{J}$:
\begin{eqnarray}\label{eq:jac}
\mathcal{J}&=&Det\left(\begin{array}{cc} \frac{\partial Q^2}{\partial \cos(\Theta)} & \frac{\partial Q^2}{\partial q^0}\\ 
\frac{\partial W}{\partial \cos(\Theta)} & \frac{\partial W}{\partial q^0} \end{array}\right)
\end{eqnarray}
whose explicit form is complicated because the invariant mass $W$ depends both on the energy transfer $q^0$ and the lepton scattering angle $\Theta$.
\section{Results of the analysis}

\label{sec:statistical}

Our main goal is to have a reliable model of weak pion production. Because the neutrino SPP data are sufficient
only to obtain information about leading axial coupling of the $\Delta(1232)$ resonance, we assume that
the extraction of vector and axial form factors can be done independently using first respective electron scattering 
and then neutrino SPP data. In the next
paragraphs we describe details of our procedure.

\subsection{Vector Contribution to Weak SPP and Electroproduction}
\label{sec:fitosip}

Our aim is (due to a poor quality of the neutrino SPP data) to reproduce correctly only
the most important characteristics of the neutrino SPP reactions. These include overall cross sections
and distributions in $Q^2$.
Detailed analysis of the electroproduction data bases on pion angular distributions. Such a task is
beyond the scope of this paper.

We use the information contained in  electron-proton $F_2$ data from \cite{Osipenko:2003ua}.  
We include 37 separate series of $F_2$ data points from the lowest value of $Q^2$ (0.225 $\mathrm{GeV^2}$) up to
2.025 $\mathrm{GeV^2}$. This $Q^2$ range overlaps with the one in ANL data.

The data desribe the inclusive structure function, thus we limit our fit to values of invariant mass $W$ up to $M_p+2m_\pi$. 
Beyond that value the experimental data contain two pion production and then more inelastic channels. With this limitation for $Q^2\leq 2.025$ and $W<M_p+2m_\pi$
we are  still left with $603$ data points.

In order to ensure that the results will reproduce well the data at the $\Delta(1232)$ peak we expanded our fit to higher value of invariant mass $W=1.27$~GeV.
There are no exclusive electron SPP data in the region $W\in(M+2m_\pi ,1.27$~GeV). Thus we 
chose to add to our fit a term in which MAID 2007 model predictions are taken as $228$ fake data points with 
errors identical to respective Osipenko \textit{et al.} \cite{Osipenko:2003ua} points.
We could not apply the MAID model directly in our fits since the exact formulas
for their SPP amplitudes have never been published. These additional
points have been generated using the on-line version of MAID (\url{http://wwwkph.kph.uni-mainz.de/MAID//}).
We have also included the information about MAID 2007 model helicity amplitudes.
The caveat is that the pion electroproduction experimental results
contain both resonant and nonresonant contributions (see e.g. Ref. \cite{Davidson:1991xz} ).
Thus the extracted helicity amplitudes depend on how one defines
the "Delta" and "background".
The HNV model differs with MAID in the treatment of both. One can not expect the resulting helicity amplitudes to be the same. From that reason the information
about helicity amplitudes has been given a large \textit{ad hoc} error assumption in our estimator.

\subsubsection{Results}

The best fit results of our vector form factor parametrization given by Eqs. (\ref{eq:modela}-\ref{eq:enda}) and Eqs. (\ref{eq:Kellyc5}-\ref{eq:c5K}) are shown in Table \ref{tab:ourfitt}. We also present there the values from Ref. \cite{Lalakulich:2006sw} in order to compare directly with our model I.
In both models the best fit value of $C_3^V(0)$ is close to the one from Ref. \cite{Lalakulich:2006sw} and we get a clear beyond-dipole $Q^2$ dependence of
$C_3^V(Q^2)$ and $C_4^V(Q^2)$. Surprisingly, the $Q^2$ dependence of $C_5^V(Q^2)$ is exactly dipole $(1+Q^2/M_V^2)^{-2}$ in model II. 
Most importantly, we have shown, that extracted form factors are model-dependent. This clearly follows from the difference of best-fit parameters between our model I and their counterpart from Ref. \cite{Lalakulich:2006sw}. One can see that, besides
the similarity in the leading form factor value $C_3^V(0)$, both fits differ by large.\\

\begin{table}[htpb]
\centering
\caption{Best fit coefficients for vector form factors given by Eqs. (\ref{eq:modela}-\ref{eq:enda}) (``Model I'') and Eqs. (\ref{eq:Kellyc5}-\ref{eq:c5K}) (``Model II'').
We do not report $1 \sigma$ errors because of hybrid character of our estimator, see explanations in the text.}
\begin{tabular}{|c|ccccccccc|}
\hline
&$C_3^V(0)$ & $C_4^V(0)$ & $A$ & $B$ & $C$ & $K_1$ & $K_2$ & $C_5^V(0)$ & $D$\\\hline
Ref. \cite{Lalakulich:2006sw}& 2.13 & -1.51 & 0.25 & 1.289 & - & - & - & 0.48 & 1.00\\
MODEL I& 2.00 & -6.77 & 0.68 & 1.40 & - & - & - & 5.95 & 1.15\\\hline
MODEL II& 2.10 & - & 4.73 & -0.39 & 5.59 & 0.13 & 1.68 & 0.62 & 1.00\\\hline
\end{tabular}
\label{tab:ourfitt}
\end{table}

\begin{figure}[htb]
\centering\includegraphics[width=\textwidth]{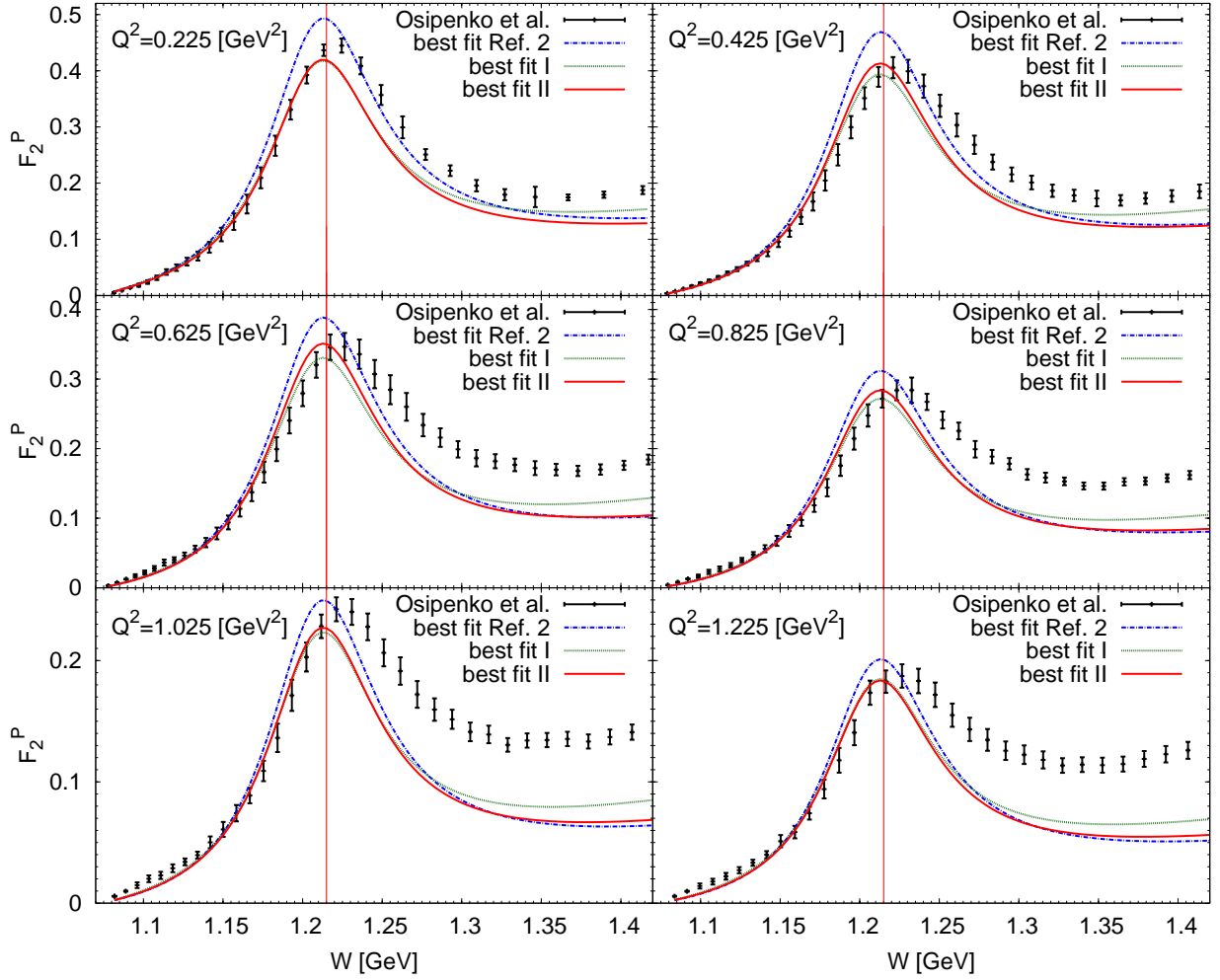}
\caption{(Color online) Best fit results for vector form factors given by Eqs. (\ref{eq:modela}-\ref{eq:enda}) (``Model I'') and Eqs. (\ref{eq:Kellyc5}-\ref{eq:c5K}) (``Model II'') plotted against experimental data
from Ref. \cite{Osipenko:2003ua} and predictions of HNV model with original Lalakulich-Paschos form factors of Ref. \cite{Lalakulich:2006sw}.
Vertical lines show the $2\pi$ production threshold.}\label{fig:ourfit1}
\end{figure}

\begin{figure}[htb]
\centering\includegraphics[width=\columnwidth]{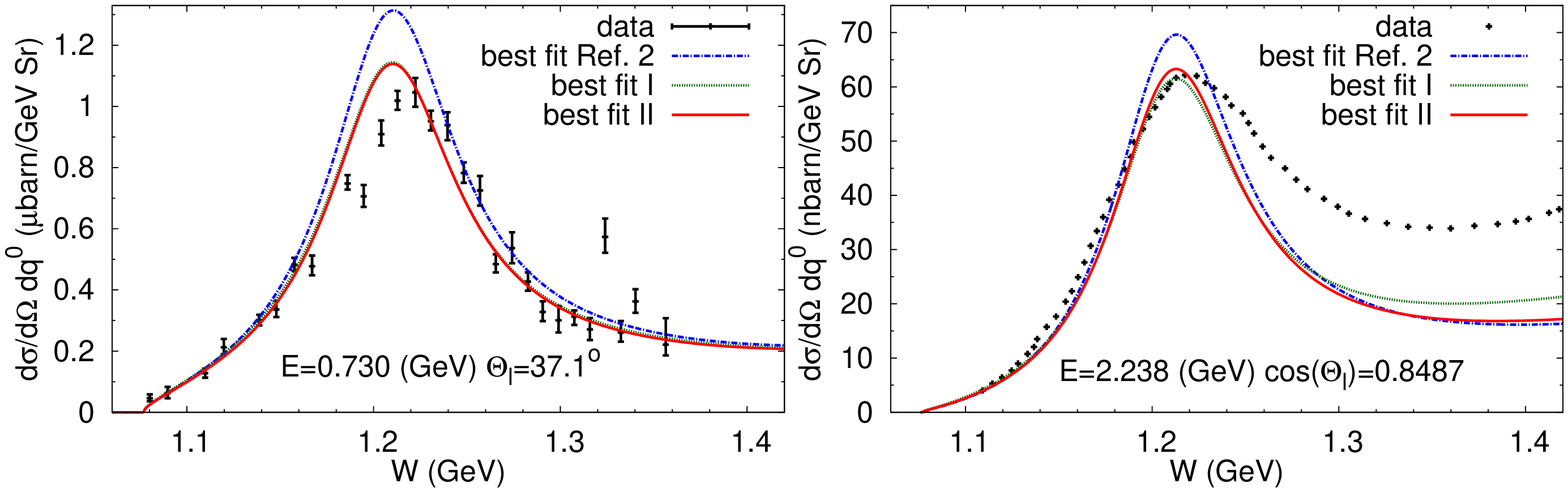}\caption{(Color online) Comparison of our best fit results and HNV model with
Lalakulich-Paschos form factors of Ref. \cite{Lalakulich:2006sw} plotted against inclusive $p(e,e')$ data (not included in the fit) from Ref.
\cite{O'Connell:1984nw} (left panel) and Ref. \cite{Christy2007} (right panel). The $Q^2$ values at peak are from left to right:
$0.1\, \mathrm{(GeV^2)}$ and $0.95\, \mathrm{(GeV^2)}$ respectively.} \label{fig:elprod}
\end{figure}

Fig. \ref{fig:ourfit1} shows that qualitatively in the
region below two pion production threshold our fit reproduces the data rather well.  
Our form factors lead to better agreement with the $F_2^p$ electron scattering data than the form factors considered in Ref. \cite{Lalakulich:2006sw}.
The same trend is clearly seen in Fig. \ref{fig:elprod}, where our best fit results are compared to the inclusive electron-proton scattering cross section data.
Inspection of Fig. \ref{fig:ourfit1} shows that biggest disagreement with data is exhibited in region of low $W$.
Our fits are going to be used in the analysis of
neutrino scattering data and some discrepancy at low $W$ is of no practical importance.

Another conclusion is that each physical model of single pion production needs its own separate resonance form factor analysis.
Any change of description of one of the elements such as $\Delta$ propagator and width, background amplitudes, unitarity constraint etc. will affect
the results. In other words, the HNV model should be used together with vector form factors fited using the HNV model in order to increase
the accuracy of its predictions.

Because both proposed form factor sets lead to very similar results, we choose to use ``Model II'' in the axial fits.


\subsection{Axial Contribution to weak SPP: fits to  bubble chamber data}
\label{sec:deuter}

We consider a statistical framework, proposed in Ref. \cite{Graczyk:2009qm}.
ANL used a neutrino beam with mean energy below 1~GeV and a large flux normalization uncertainty $\Delta p_{ANL}\sim 20\%$
that was not included in the published $d\sigma/dQ^2$ cross section for the reaction in Eq. (\ref{channel_1}) \cite{Graczyk:2009qm}.
ANL reported the data with the invariant mass cut $W<1.4$ GeV, which allows us
to confine to the $\Delta(1232)$ region. We can neglect contributions from heavier resonances, whose axial couplings are by large unknown.
Our analysis uses information from all available SPP isospin channels. The detiled description of the statistical approach can be found in Ref. \cite{Graczyk:2014dpa}.

We treat $C_5^A(0)$, $M_{A\Delta}$ and 
normalization factor $p_{ANL}$ as free fit parameters in the analysis of axial $N\to\Delta$(1232) transition. We present our results in Tab. \ref{tab:axfit}.
\begin{table}[htpb]
\caption{Best fit for the $\Delta$(1232) axial form factors on deuteron target. 
  Errors for $C_5^A(0)$ and $M_{A\Delta}$ were obtained after marginalization of $p_{ANL}$. }
\label{tab:axfit}\centering
\begin{tabular}{|c|ccccc|}
\hline
Channel & $C_5^A(0)$ & $M_{A\Delta}$(GeV) & $p_{ANL}$ & $\chi^2/NDF$ & $NDF$ \\
\hline
$\nu_\mu+p\to\mu^-+p+\pi^+$ & 1.11$\foot{+0.32}{-0.34}$ & 0.97$\foot{+0.17}{-0.17}$ & 1.04 & 0.20 &6 \\
$\nu_\mu+n\to\mu^-+p+\pi^0$ & 1.31$\foot{+0.49}{-0.77}$ & 1.00$\foot{+0.27}{-0.25}$ & 0.93 & 1.52 & 9 \\
$\nu_\mu+n\to\mu^-+n+\pi^+$ & 2.83$\foot{+0.62}{-0.60}$ & 0.76$\foot{+0.13}{-0.13}$ & 0.94 & 1.47 & 9 \\
Joint fit & 1.10$\foot{+0.15}{-0.14}$ & 0.85$\foot{+0.09}{-0.08}$ & 0.90 & 2.06& 30 \\
\hline
\end{tabular}
\end{table}
where fits to all three channels separately as well as the joint fit to three channels are listed.
In each case the number of degree of freedom is calculated as:\\ $NDF=$ No. $Q^2$ bins $-$ No. fitted parameters.\\
We see, that taken separately the $p\pi^+$ (A1) and $p\pi^0$ (A2) channels are statistically
consistent, albeit their
predicted scale parameters differ by around 10\%. The latter channel seems to carry less information on the $N\to \Delta$ 
transition axial current
than the first one. This fact is reflected in larger uncertainties. We explain it by a bigger background contribution to that channel,
which makes it less sensitive to changes in the
$\Delta$ resonance form factors.

The $n\pi^+$ (A3) channel gives results inconsistent with the other two. $C_5^A(0)$ is obtained twice as large
as for the $p\pi^+$ and $p\pi^0$ channels and $M_{A\Delta}$ significantly smaller. Here the number of events reported by ANL is comparable
to $p\pi^0$ channel, but theoretical cross section predicted by our model are smaller. This results in the overestimation of $C_5^A(0)$.
Surprisingly, the fits to separate isospin channels give acceptable values of $\chi^2_{min}$ for both  neutron channels.

Deuteron effects affect mostly the value of $C_5^A(0)$ by up to 20\%. In the joint fit on free proton and neutron targets
we obtained $C_5^A(0)=0.93\foot{+0.13}{-0.13}$ and $M_{A\Delta}=0.81\foot{+0.09}{-0.09}\;$GeV compared to $1.10$ and $0.85\;$GeV in the deuteron case. A significant improvement with respect to previous fits to HNV model done in
Refs. \cite{Hernandez:2007qq,Hernandez:2010bx}
is that with deuteron target effects we get the best fit value of $C_5^A(0)$ within 1$\sigma$ range from the
theoretical Goldberger-Treiman relation.
The joint fit agrees also on the 1$\sigma$ level with separate fits on $p\pi^0$ and $p\pi^+$ channels.

\begin{figure}[htb]
\centering\includegraphics[width=\columnwidth]{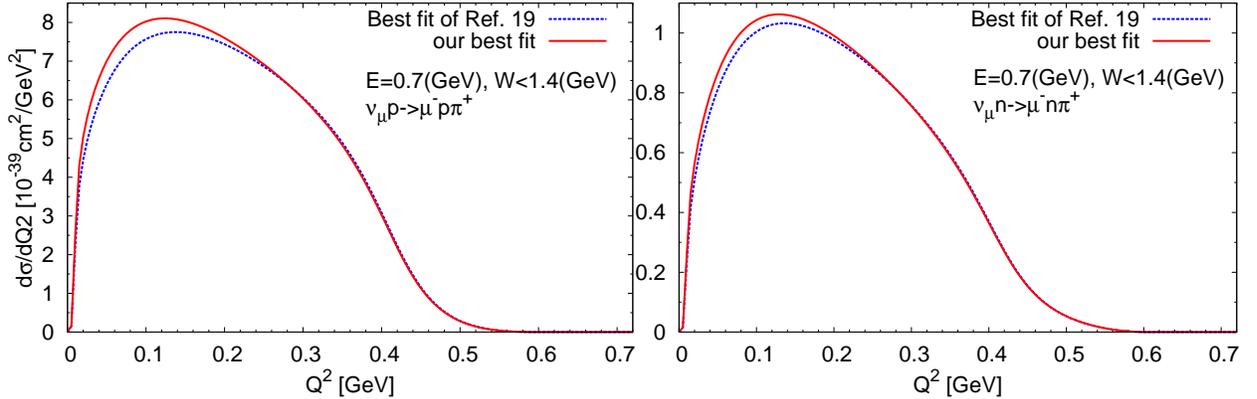}\caption{(Color online) Differential cross sections
on a free nucleon target for our best fit form factors and for fits of Ref. \cite{Hernandez:2010bx}.} \label{fig:dq2}
\end{figure}
\begin{figure}[htb]
\centering\includegraphics[width=\columnwidth]{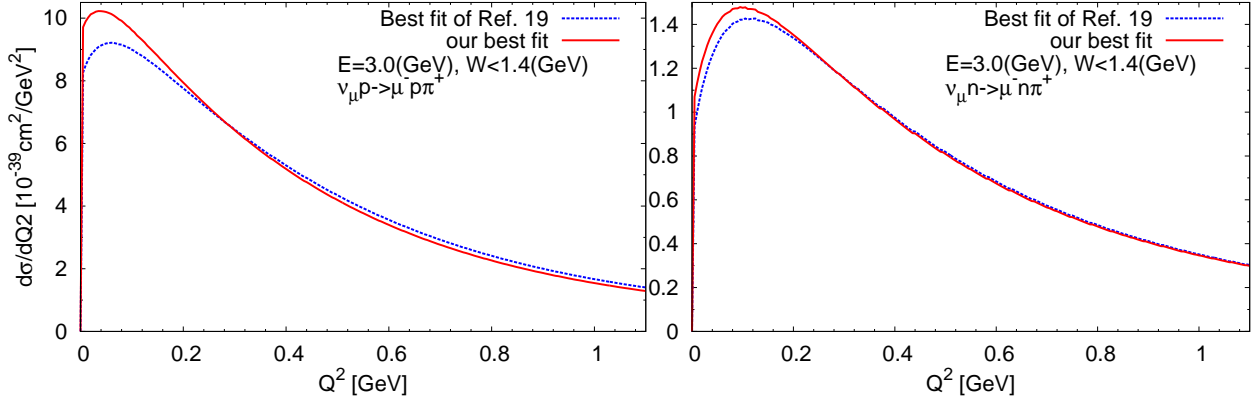}\caption{(Color online) Same as Fig. \ref{fig:dq2} but for $E_\nu=3\;$GeV.} \label{fig:dq23}
\end{figure}
We have compared our best fit form factors with the results of Ref. \cite{Hernandez:2010bx} for two different neutrino energies characteristic for Minos and MINERvA
experiments in Figs. \ref{fig:dq2} and Fig. \ref{fig:dq23}. Our form factors predict different shape and magnitude of $d\sigma/dQ^2$.
The size of effect is bigger for the lower values uf $Q^2$ and for proton channel, where the $\Delta$ contribution dominates. In the latter
one has up to 10\% difference in the cross section.

Normalization factors $p_{ANL}$ fitted separately for each channel are different for neutrons and protons.
The proton channel prefers the data to be scaled up and both neutron channels 
prefer the data to be scaled down. The joint fit uses the same $p_{ANL}$ parameter
for all channels and seems to prefer the data to be scaled down even more ($p_{ANL}\approx 0.90$).
The values of $p_{ANL}$ are all within the assumed flux normalizadion error $\Delta p_{ANL}$.

Finally, we noticed that the best fit values for $C_5^A(0)$ and $M_{A\Delta}$ in the $p\pi^+$ (A1) channel
are different from those obtained in Ref. \cite{Graczyk:2009qm}. We explain it by the inclusion of nonrenonant background 
in the current analysis.

\section{Conclusions}
\label{sec:conclusions}

In this paper we review a new attempt to get an information about weak $N\to\Delta$ transition
matrix elements previously presented in Ref. \cite{Graczyk:2014dpa} as well as we extent this analysis by presenting the newest fits of the vector form factors and discussion of the axial form factors. The fit to electromagnetic $F_2^p$ has clearly shown, that the extracted vector form factors
of the $\Delta$ resonance are model-dependent, \eg the HNV model gives the best results with the form factors extracted using the full HNV model, as it has been done in this paper.

We discussed axial form factor fits obtained based on the analysis of all three neutrino SPP channels, including the neutrino-neutron channels. In previous works usually only neutrino-proton channel was utilized to extract the axial form factors. A critical analysis of neutrino-neutron channel, on qualitative level, appears also in other papers  see \eg \cite{Fogli:1979cz,Hernandez:2007qq}, a detailed discussion can be found in Ref. \cite{Lalakulich:2010ss} as well. The obtained value of $C_5^A(0)$ agrees, on the 1$\sigma$ level,
with the Goldberger-Treiman relation if the deuteron effects are taken into account, which is an important result. If one neglects the nuclear effects  the resulting $C_5^A(0)$ value is lower.
Also, there is a strong tension between $n\pi^+$ and remaining two channels (see also \cite{Hernandez:2007qq} and \cite{Lalakulich:2010ss}). The same theoretical
model does not seem to give a consistent reproduction of data in all channels.

There can be various reasons for that. Firstly the existing bubble chamber data on neutron SPP channels are of poor statistics.
Secondly, the chiral model for the background is well justified only near the pion production threshold. It may be not
 reliable in the $\Delta(1232)$ peak region. Last, but not least, the $n\pi^+$ channel is subject to large FSI effects,
 as shown in Ref. \cite{Sato_FSI}. Thus the spectator model used both in experimental analyses of
 ANL and BNL as well as in our calculations may give invalid results in this channel. The plots in
 Ref. \cite{Sato_FSI} suggest a reduction of the $n\pi^+$ cross section due to FSI. Further studies are needed.

Still another reason of these difficulties may come from a missing unitarization of the model. This constraint, following
the Watson theorem \cite{Watson:1952ji}, imposes a relation between phases in lepton-nucleon and pion-nucleon elastic scattering amplitudes. Unitarity is not satisfied in our approach. 
In a recent study Nieves, Alvarez-Ruso, Hernandez and Vicente-Vacas \cite{nieves_nuint14} tried to correct the HNV model by introducing 
phenomenological phases. They obtained better agreement of the best fit value of $C_5^A(0)$ with the Goldberger-Treiman relation\footnote{See also the contribution of E. Hernandez presented during the CETUP* 2014 workshop.}. This is a strong indication of the importance of
proper model unitarization for the pion neutrinoproduction case. More studies are necessary.

Relatively large values of experimental errors and our inability to extract independent $C_3^A$ and $C_4^A$ form factors implies, that better statistics SPP measurements in the $\Delta$ region on proton or deuteron targets are badly needed. Keeping
in mind difficulties in the treatment of nuclear effects on heavier targets it is the only way to get precise information about the
$N\to \Delta$ axial transition matrix elements. 


\begin{theacknowledgments}
We thank  U. Mosel for his remarks  on the previous version of the paper. JZ was supported by Grant UMO-2011/M/ST2/02578. Numerical calculations were carried out in Wroclaw Centre for Networking and Supercomputing 
(\url{http://www.wcss.wroc.pl}),
grant No. 268.
\end{theacknowledgments}



\bibliographystyle{aipproc}   

\bibliography{bibdratold}

\IfFileExists{\jobname.bbl}{}
 {\typeout{}
  \typeout{******************************************}
  \typeout{** Please run "bibtex \jobname" to optain}
  \typeout{** the bibliography and then re-run LaTeX}
  \typeout{** twice to fix the references!}
  \typeout{******************************************}
  \typeout{}
 }

\end{document}